\documentclass[%
 reprint,
 amsmath,amssymb,
 aps,
]{revtex4-2}

\usepackage{graphicx}
\usepackage{dcolumn}
\usepackage{bm}
\usepackage{amsmath}
\usepackage{color}

\usepackage[caption=false]{subfig}
\begin{document}

\preprint{APS/123-QED}

\title{Intermittent random walks under stochastic resetting}

\author{Rosa Flaquer-Galmés}
\affiliation{Grup de Física Estadística. Departament de F\'{\i}sica, Universitat Aut\`{o}noma de Barcelona, 08193 Bellaterra, Spain}

\author{Daniel Campos}
\affiliation{Grup de Física Estadística. Departament de F\'{\i}sica, Universitat Aut\`{o}noma de Barcelona, 08193 Bellaterra, Spain}

\author{Vicen\c c M\'endez}
\affiliation{Grup de Física Estadística. Departament de F\'{\i}sica, Universitat Aut\`{o}noma de Barcelona, 08193 Bellaterra, Spain}

\begin{abstract}
We analyze a one-dimensional intermittent random walk on an unbounded domain in the presence of stochastic resetting. In this process, the walker alternates between local intensive search, diffusion, and rapid ballistic relocations in which it does not react to the target. We demonstrate that Poissonian resetting leads to the existence of a non-equilibrium steady state. We calculate the distribution of the first arrival time to a target along with its mean and show the existence of an optimal reset rate. In particular, we prove that the initial condition of the walker, i.e., either starting diffusely or relocating, can significantly affect the long-time properties of the search process. Moreover, we demonstrate the presence of distinct parameter regimes for the global optimization of the mean first arrival time when ballistic and diffusive movements are in direct competition.
\end{abstract}

\maketitle


\section{Introduction}

In recent years, significant efforts have been dedicated to studying optimal search strategies \cite{StKr86} in the field of behavioral ecology, both theoretically and experimentally \cite{KrMc15,MeCaBa16,Ka12,Vi11}. A paradigmatic kind of model often invoked to capture the characteristics of animal trajectories is that of intermittent movement, where different motion patterns or mechanisms are alternated in time by the organism \cite{Os09,Re06,bo14}. One such of these intermittent search strategies is the Intermittent Random Walk (IRW) which combines two distinct phases. The first one (known as 'extensive' search) consists of rapid ballistic relocations that are often unresponsive to the presence of the target, as it is considered that these movements are used by the animals just for choosing a new region to explore, and cognitive abilities are partially or completely suppressed during this transit \cite{Benichou_2011}. This relocation is then followed by a slower diffusive phase (often denoted as 'intensive local' search) during which the target can be in general detected if approached \cite{Benichou_2011}. 

IRWs represent then a natural and significant extension of Brownian trajectories in many contexts \cite{zhou2020}. In biological systems, they have been tested as plausible mechanisms for seed dispersal \cite{abdullahi2019} or copepod dispersal \cite{schmitt2001}. Their use in physical systems, however, is also widespread. They have been proved, for instance, to offer a plausible framework to describe particle trajectories at liquid-solid interfaces \cite{skaug2013}, supercooled liquids \cite{cavagna2009}, colloidal suspensions \cite{weeks2002}, or in different types of disordered media \cite{rajasekar1998,deanna2013,kang2014}, among many other \cite{pastore2017}. Also, their convenience as a mechanism to optimize energy minimization algorithms have also been explored \cite{chow2015}. Finally, their formal connection to self-trapping and/or self-avoiding interactions has also been suggested as a possible underlying mechanism to explain such trajectories at microscopic scales \cite{grassberger2017}.

Theoretical studies have demonstrated that IRWs are also an effective search strategy in a wide range of scenarios \cite{BeCo05,ramezanpour2007,Os07,Lo08,oshanin2009,Rojo2010,GoCaMe11}. For the case of Poissonian switching between the relocation and diffusion modes, an optimal relationship between the two switching rates can minimize the mean search time for detecting a randomly localized target in bounded domains \cite{BeCo05}. Despite the potential interest of this model, a handicap of IRWs when regarded as search trajectories is that their mean search time is still infinite in unbounded domains, as happens with Brownian walkers and other strategies totally or partially controlled by a homogeneous Gaussian noise. As a remedy to this, the mechanism of stochastic resetting (in which the walker returns to its initial position after a random sojourn time, and so arbitrary departures from the target are avoided) has been advocated. In the particular case of biological foraging, stochastic resetting can additionally offer a new layer of realism to the mechanism of search. Indeed, many organisms repeatedly return to a central nest, or to previously visited sites, as a part of their regular foraging strategies \cite{O23b}. This revisiting process can then be effectively described by introducing a stochastic resetting of the individual/animal position to those preferential points. In such cases, stochastic resetting is capable of turning the mean first passage finite. Furthermore, there exists an optimal value for the rate at which stochastic resets are executed such that the corresponding mean first passage time attains its minimum value.\cite{EvMa11}.

In this work, we explore the main transport properties of an IRW model combining diffusive and relocation modes and study its dynamics under stochastic Poissonian resetting. This will allow us to study whether the existence of an unresponsive mode (with relocations during which the target cannot be detected) modifies significantly the arrival statistics to a target. In Section II we review the properties of the mean square displacement of the walker, and the first arrival time to a hidden target in the absence of resetting. After that, in Section III we compute the non-equilibrium stationary state reached by the system when resetting is introduced, how the statistics of the first arrival probability get modified accordingly, and its dependence on the initial condition chosen. An exhaustive analysis of the optimal mean first arrival time in terms of the characteristic parameters is performed. It reveals an interesting trade-off of the switching rates between the relocation and diffusive states. Finally, we summarize in Section IV the main conclusions of our work.

\section{IRW model} 

Let $\rho_{+}(x,t)$ and $\rho_{-}(x,t)$ be the probability density of a walker at position $x$ at time $t$ with velocities $+v$ and $-v$, respectively. States $+$ and $-$ correspond to a relocation mode in which the walker moves ballistically during a random (and exponentially distributed) time before switching to the diffusive mode. After that, the walker moves diffusively for a random time (in which it is also able to detect the target), also exponentially distributed, before switching back to the relocation mode. If the probability density for the walker to be at position $x$ at time $t$ moving diffusively is $\rho_{0}(x,t)$, then

\begin{align}
\frac{\partial\rho_{+}}{\partial t}&=-v\frac{\partial\rho_{+}}{\partial x}-\frac{1}{\tau_v}\rho_{+}+\frac{1}{2\tau_D}\rho_{0},\label{eq:m3e1} \\
\frac{\partial\rho_{-}}{\partial t}&=v\frac{\partial\rho_{-}}{\partial x}-\frac{1}{\tau_v}\rho_{-}+\frac{1}{2\tau_D}\rho_{0},\label{eq:m3e2} \\
\frac{\partial\rho_{0}}{\partial t}&=D\frac{\partial^{2}\rho_{0}}{\partial x^{2}}+\frac{1}{\tau_v}\big(\rho_{+}+\rho_{-}\big)-\frac{1}{\tau_D}\rho_{0},\label{eq:m3e3}
\end{align}
where $\tau_D^{-1}$ and $\tau_v^{-1}$ are the switching rates from diffusive to relocation mode, and from relocation to diffusive mode, respectively. Note that the factor $1/2$ in the last term of the right-hand side of Eqs. \eqref{eq:m3e1} and \eqref{eq:m3e2} implies that when switching from diffusion to relocation, right and left moving states are equally probable as there is no predilect direction. Nevertheless, for systems where the left-right symmetry is broken in the relocation phase, one can generalize Eqs. \eqref{eq:m3e1} to \eqref{eq:m3e3} by introducing the corresponding probability, $p$ to relocate to the right and $1-p$ to relocate to the left. Let us introduce now the Fourier transform (denoted with a bar symbol) and the Fourier-Laplace transform (denoted with a hat symbol) as follows,

\begin{align*}
       &\bar{\rho_i}(k,t)=\int_{-\infty}^{\infty}e^{-ikx}\rho_i (x,t)dx,\;\; \\ &\hat{\rho_i}(k,s)=\int_{-\infty}^{\infty}e^{-ikx}\int_{0}^{\infty}e^{-st}\rho_i (x,t)dxdt 
\end{align*}
with $i=0,+,-$. Transforming Eqs. (\ref{eq:m3e1}-\ref{eq:m3e3}) by Fourier-Laplace we can write a single equation for $\hat{\rho}_0(k,s)$:

\begin{align}
        s\hat{\rho}_{0}&(k,s)-\bar{\rho}_{0}(k,0) = -Dk^{2}\hat{\rho}_{0}(k,s) \nonumber\\ 
    & + \frac{1}{\tau_D}\left[\frac{(s\tau_v + 1)-(s\tau_v + 1)^{2}-(kv\tau_v)^{2}}{(s\tau_v + 1)^{2}+(kv\tau_v)^{2}}\right]\hat{\rho}_{0}(k,s)  \nonumber\\  
    & + \frac{(s\tau_v + 1)\left(\bar{\rho}_{+}(k,0) + \bar{\rho}_{-}(k,0) \right)}{(s\tau_v + 1)^{2}+(kv\tau_v)^{2}}   \nonumber\\
     &+ ikv\tau_v \frac{\left(\bar{\rho}_{+}(k,0) - \bar{\rho}_{-}(k,0) \right)}{(s\tau_v + 1)^{2}+(kv\tau_v)^{2}},
\label{sr}
\end{align}
where $\bar{\rho_i}(k,0)$ is the Fourier transform of the initial condition at state $i$. We study two different initial conditions for a walker located at $x=x_0$ at time $t=0$. For the first case (case I in the following) we consider the walker being at the diffusive state $i=0$, so that  $\bar{\rho}_{0}(k,0) = e^{-ikx_0}$ and $\bar{\rho}_{+}(k,0) = \bar{\rho}_{-}(k,0) = 0$. For the second one (case II), we consider a walker in the relocation state with the same probability of being at states $i=+,-$, so that $\bar{\rho}_0(k,0)=0$ and $\bar{\rho}_+(k,0)=\bar{\rho}_-(k,0)=e^{-ikx_0}/2$. To facilitate reading of the manuscript, we will use superscripts $^{D}$ and $^{v}$ (referring to the initial state, diffusive or ballistic, in each case) to identify our results for cases I and II.  One could also consider that after the resetting the system is in a random phase, that is, it could be diffusing or relocating with the same probability after resetting. This third case is included in Appendix \ref{app:randomIC} as the results obtained for it can be understood via the first two cases here presented. The distinction between these initial conditions is relevant as in Section \ref{secResetting} we show that in the presence of resetting the effect of the initial condition cannot be neglected. For case I the propagator $\hat{\rho}(k,s)=\hat{\rho}_0(k,s)+\hat{\rho}_{+}(k,s)+\hat{\rho}_{-}(k,s)$ reads:

\begin{align}
    \hat{\rho}^D(k,s) = \hat{\rho}_0^{D}(k,s) \left[ 1 + \frac{\tau_v}{\tau_D}\frac{s\tau_v + 1}{(s\tau_v + 1)^2 + (v\tau_vk)^2} \right],
    \label{wholeD}
\end{align}
where $\hat{\rho}^D_{0}(k,s)$ is defined as

\begin{align}
    \hat{\rho}_0^{D}(k,s) = \frac{e^{-ikx_0}}{s + Dk^2 + \frac{1}{\tau_D}\big[ 1 - \frac{s\tau_v + 1}{(s\tau_v + 1)^2 + (vk\tau_v)^2} \big]} . \label{D}
\end{align}
Analogously, for case II these expressions are

\begin{align}
    \hat{\rho}^v(k,s) = & \hat{\rho}_0^{v}(k,s) \left[ 1 + \frac{\tau_v}{\tau_D}\frac{s\tau_v + 1}{(s\tau_v + 1)^2 + (v\tau_vk)^2} \right] \nonumber \\
    & + \tau_v e^{-ikx_0}\frac{s\tau_v + 1}{(s\tau_v + 1)^2 + (v\tau_vk)^2},
    \label{wholev}
\end{align}

\begin{align}
    \hat{\rho}_0^{v}(k,s) = \hat{\rho}_0^{D}(k,s)\frac{(s\tau_v + 1)}{(s\tau_v + 1)^2 + (vk\tau_v)^2}.
    \label{v}
\end{align}

\begin{figure}[h!]
    \centering
    \includegraphics[width=\linewidth]{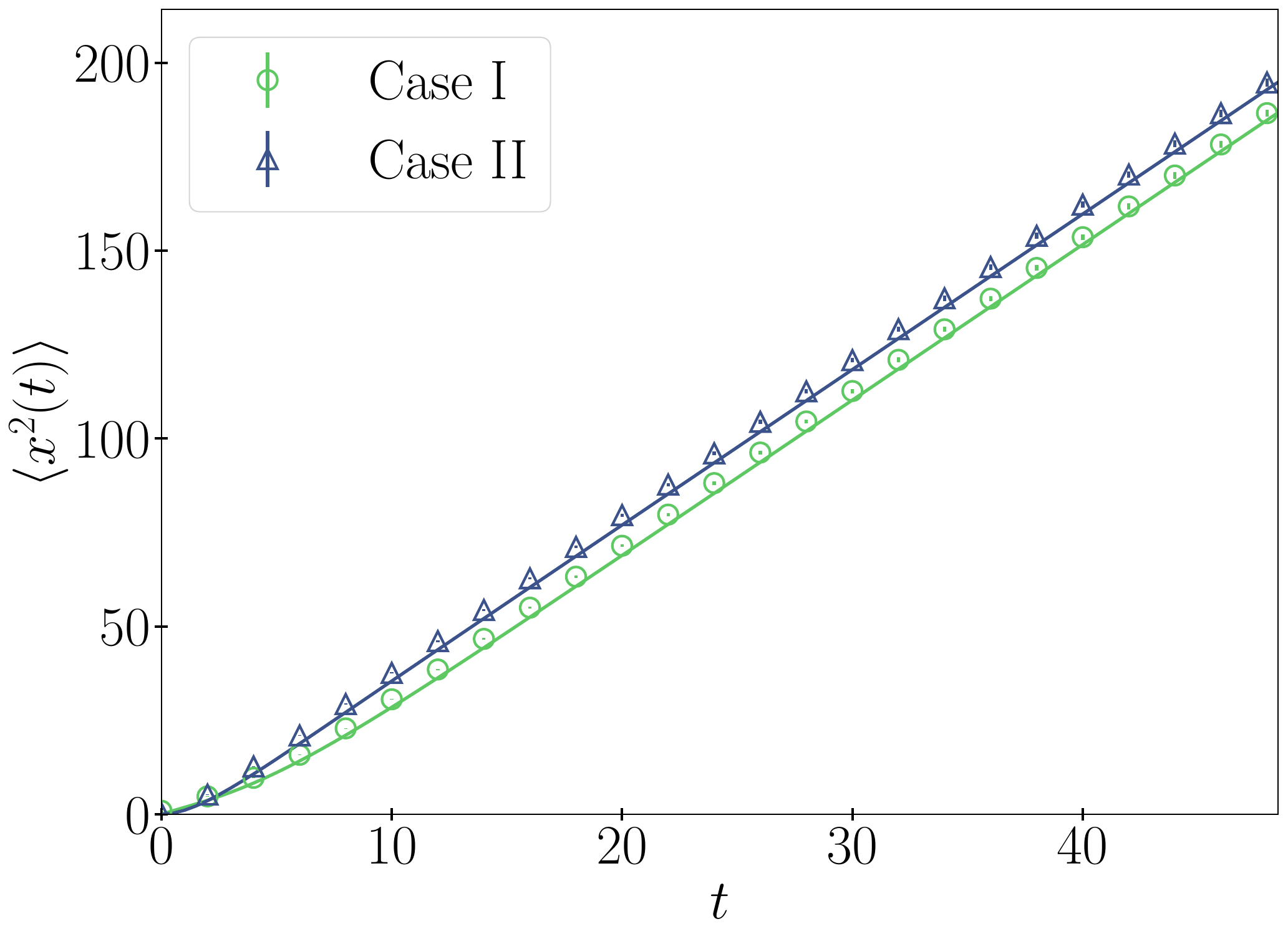}
    \caption{Mean squared displacement for case I (green circles) and case II (blue triangles). In solid lines we show the analytic result, Eq. \eqref{msdsr}, and in dots data obtained from numerical simulations. The parameters used are $D=v=1$, $\tau_D = 3.8$, $\tau_v = 3.3$ and $x_0=0$.}
    \label{fig:msdsr}
\end{figure}

To study the transport properties of this IRW, we start by computing the Mean Squared Displacement (MSD) of the whole process. We also note that, due to the symmetry of the initial conditions, the mean distance traveled is 
$\left\langle x(s)\right\rangle =\frac{1}{i}\left[\frac{\partial\hat{\rho}(k,s)}{\partial k}\right]_{k=0}=0,$
where $\hat{\rho}(k,s)$ is the propagator of the whole process (Eq. \eqref{wholeD} or \eqref{wholev} depending on the initial condition). Then, the MSD can be computed from
$
\left\langle x^{2}(s)\right\rangle =-\left[\frac{\partial^{2}\hat{\rho}(k,s)}{\partial k^{2}}\right]_{k=0}.
$
Assuming $x_0=0$ for simplicity, when transforming back from Laplace space the MSD reads
\begin{align}
    \left\langle x^{2}(t)\right\rangle &=\frac{2(D+\frac{(v\tau_v)^{2}}{\tau_D})}{(1+\frac{\tau_v}{\tau_D})}t+A_{1} \nonumber \\ 
    & +A_{2}e^{-\frac{t}{\tau_v}}+A_{3}e^{-(\frac{1}{\tau_v}+\frac{1}{\tau_D})t},
\label{msdsr}
\end{align}
where the coefficients $A_i$ with $i=1,2,3$ take different values depending on whether the walker is initially in the diffusive or relocation state; specific values of $A_i$ are included in Appendix \ref{app: Coefficients MSD}. Note that in the large time limit, regardless of the initial condition, the MSD is essentially diffusive and behaves according to
\begin{align}
    \left\langle x^{2}(t)\right\rangle =2\frac{D\tau_D+(\tau_vv)^{2}}{\tau_D+\tau_v}t\quad\textrm{as\ensuremath{\quad t\rightarrow\infty}}
\end{align}
On the other hand, for short times one recovers a diffusive MSD for case I, while for case II is ballistic, as expected given the initial conditions:

\begin{align}
    \left\langle x^{2}(t)\right\rangle^D& \approx 2Dt \quad &\mathrm{as} \quad t\to0\\
    \left\langle x^{2}(t)\right\rangle^v& \approx \left(v^2 + \frac{D}{\tau_v}\right)t^2  \quad &\mathrm{as} \quad t\to0
\end{align}

In Fig. \ref{fig:msdsr} we have compared Eq. \eqref{msdsr} for cases I and II with numerical simulations.
We conclude then that the switching processes between diffusive and relocation states on affects the transient temporal behaviour of the MSD.

\subsection{First arrival time}

A widely used measure of the efficiency of a search process is the study of the statistics of a walker arriving for the first time at a given position where a hidden target is located, that is, the First Arrival Time (FAT) when the target position is considered as an absorbing boundary. In particular, we consider the First Arrival Time Density (FATD) and its Mean (MFAT). As stated above, we work under the assumption that the target can only be detected when the walker is in the diffusive phase, as is assumed that in the relocation phase the high velocity prevents the walker from detecting the target \cite{Benichou_2011}. Considering searchers starting from an initial position $x=x_0$ and a target located at $x=0$ one can rewrite Eq. \eqref{eq:m3e3} to account for the target

\begin{align}
    \frac{\partial\rho_{0}}{\partial t}&=D\frac{\partial^{2}\rho_{0}}{\partial x^{2}}+\frac{\left(\rho_{+}+\rho_{-}\right)}{\tau_v}-\frac{\rho_{0}}{\tau_D} -f(x_0,t)\delta(x), \label{eq:m3e3abs}
\end{align}
while Eq. \eqref{eq:m3e1} and Eq. \eqref{eq:m3e2} remain unchanged. Here, $f(x_0,t)$ is the first arrival probability density to a target at the origin. The role of the last term in \eqref{eq:m3e3abs} is to remove the particle when it arrives at a target placed at $x=0$. To compute the FATD we transform Eqs. \eqref{eq:m3e1}, \eqref{eq:m3e2} and \eqref{eq:m3e3abs} by Fourier-Laplace and combine them in a single equation for $\rho_0$. Integrating this equation over $k$ and using that $\int_{-\infty}^{\infty} dk\hat{\rho}_{0}(k,s) = \bar{\rho}_{0}(x=0,s) = 0$, we obtain the FATD for case I, denoted with the superscript $D$, which reads

\begin{align}
    \tilde{f}^{D}(x_0,s)=\frac{I(x_{0},s)}{I(x_{0}=0,s)},
\end{align}
where  $\tilde{f}(x_0,s)$ is the Laplace transform of $f(x_0,t)$ and 

\begin{align}
    I(x_{0},s) &= \frac{1}{2\pi}\int_{-\infty}^{\infty} \hat{\rho}_0^{D}(k,s)dk  \nonumber \\
                &= \frac{1}{2\pi}\int_{-\infty}^{\infty}\frac{A(k,s)e^{-ikx_{0}}}{k^{4}+B(s)k^{2}+C(s)}dk.
                \label{eq:I}
\end{align}
The coefficients $A(k,s)$, $B(s)$ and $C(s)$ are defined as:

\begin{align}
A(k,s)=\frac{\left(1+s\tau_{v}\right)^{2}}{D(v\tau_{v})^{2}}+\frac{k^{2}}{D},
\label{eq:A}
\end{align}

\begin{align}
B(s)=\frac{1+s\tau_D}{D\tau_D}+\frac{\left(1+s\tau_{v}\right)^{2}}{(v\tau_{v})^{2}},
\label{eq:B}
\end{align}

\begin{align}
C(s)=\frac{\left(1+s\tau_{v}\right)\left[s\tau_{v}+s\tau_{D}\left(1+s\tau_{v}\right)\right]}{D\tau_{D}(v\tau_{v})^{2}}.
\label{eq:C}
\end{align}
The integral (\ref{eq:I}) can be rewritten in the more suitable form
\[
I(x_{0},s)=\frac{1}{2\pi}\int_{-\infty}^{\infty}\frac{A(k,s)e^{-ikx_{0}}}{\left[k^{2}+\lambda_{+}^{2}\right]\left[k^{2}+\lambda_{-}^{2}\right]}dk,
\]
with
\begin{equation}
\lambda_{\pm}=\left[\frac{B(s)\pm\sqrt{B(s)^{2}-4C(s)}}{2}\right]^{1/2}.
\label{eq:lambda}
\end{equation}
It is not difficult to show that $B^2(s)>4C(s)$ so that $\lambda_{\pm}$ are always real values.
From the residue theorem, the integral can be computed
\begin{align}
        I(x_{0},s)&=\frac{A(k=i\lambda_{-},s)e^{-|x_{0}|\lambda_{-}} }{2\lambda_{-}(\lambda_{+}^{2}-\lambda_{-}^{2})} -\frac{A(k=i\lambda_{+},s)e^{-|x_{0}|\lambda_{+}}}{2\lambda_{+}(\lambda_{+}^{2}-\lambda_{-}^{2})} \nonumber\\ 
        &=\frac{1}{2(\lambda_{+}^{2}-\lambda_{-}^{2})}\left[ \frac{\phi_{-}e^{-|x_{0}|\lambda_{-}}}{D} - \frac{\phi_{+}e^{-|x_{0}|\lambda_{+}}}{D}  \right],
\end{align}
where we have defined $\phi_{\pm}(s)$ as
\begin{equation}
\phi_{\pm}(s)=\frac{\left(1+s\tau_{v}\right)^{2}}{(v\tau_{v})^{2}\lambda_{\pm}}-\lambda_{\pm}.
\label{fis}
\end{equation}

From this, the FATD finally reads
\begin{equation}
    \tilde{f}^{D}(x_0,s)=\frac{\phi_{-}e^{-|x_0|\lambda_{-}}-\phi_{+}e^{-|x_0|\lambda_{+}}}{\phi_{-}-\phi_{+}}
    \label{eq:FATD_d}
\end{equation}
Now, for case II, denoted with the superscript $v$, the FATD in the Laplace space follows straightforwardly,
\begin{equation}
    \tilde{f}^{v}(x_0,s)=\frac{J(x_0,s)}{I(x_0=0,s)},
\end{equation}
where $J(x_0,s)$ is defined as
\begin{align}
J(x_{0},s)&=\frac{1}{2\pi}\frac{1+s\tau_v}{D(v\tau_v)^2}\int_{-\infty}^{\infty}\frac{e^{-ikx_{0}}}{k^{4}+B(s)k^{2}+C(s)}dk\nonumber\\
&=\frac{1+s\tau_v}{D(v\tau_v)^2}\frac{\lambda_{+}e^{-|x_{0}|\lambda_{-}} - \lambda_{-}e^{-|x_{0}|\lambda_{+}} }{2\left(\lambda_{+}^{2}-\lambda_{-}^{2}\right)\lambda_{-}\lambda_{+}}.
\label{eq:J(x0,s)}
\end{align}
Following the same reasoning as for case I, the final expression for the first arrival distribution reads

\begin{equation}
    \tilde{f}^{v}(x_0,s)=\frac{1+s\tau_v}{(v\tau_v)^2}\frac{\lambda_{+}e^{-|x_0|\lambda_{-}}-\lambda_{-}e^{-|x_0|\lambda_{+}}}{\lambda_{+}\lambda_{-}(\phi_{-} - \phi_{+} )}.
    \label{eq:FATD_v}
\end{equation}

If one tries to compute the MFAT as $\langle T (x_0) \rangle = \int_0^\infty t f(x_0,t)dt$ it goes to infinity as walkers are not confined. We note that this is different from the results reported previously for the IRW, e.g. in \cite{BeCo05}, \cite{Lo08} as in there the walker was considered to be in a finite domain or interval. Thus, it is clear that in unbounded domains the walker needs another mechanism apart from the IRW in order to optimize the search of a target.

\section{Resetting} \label{secResetting}

\begin{figure}[h!]
	\centering
	\includegraphics[width=\linewidth]{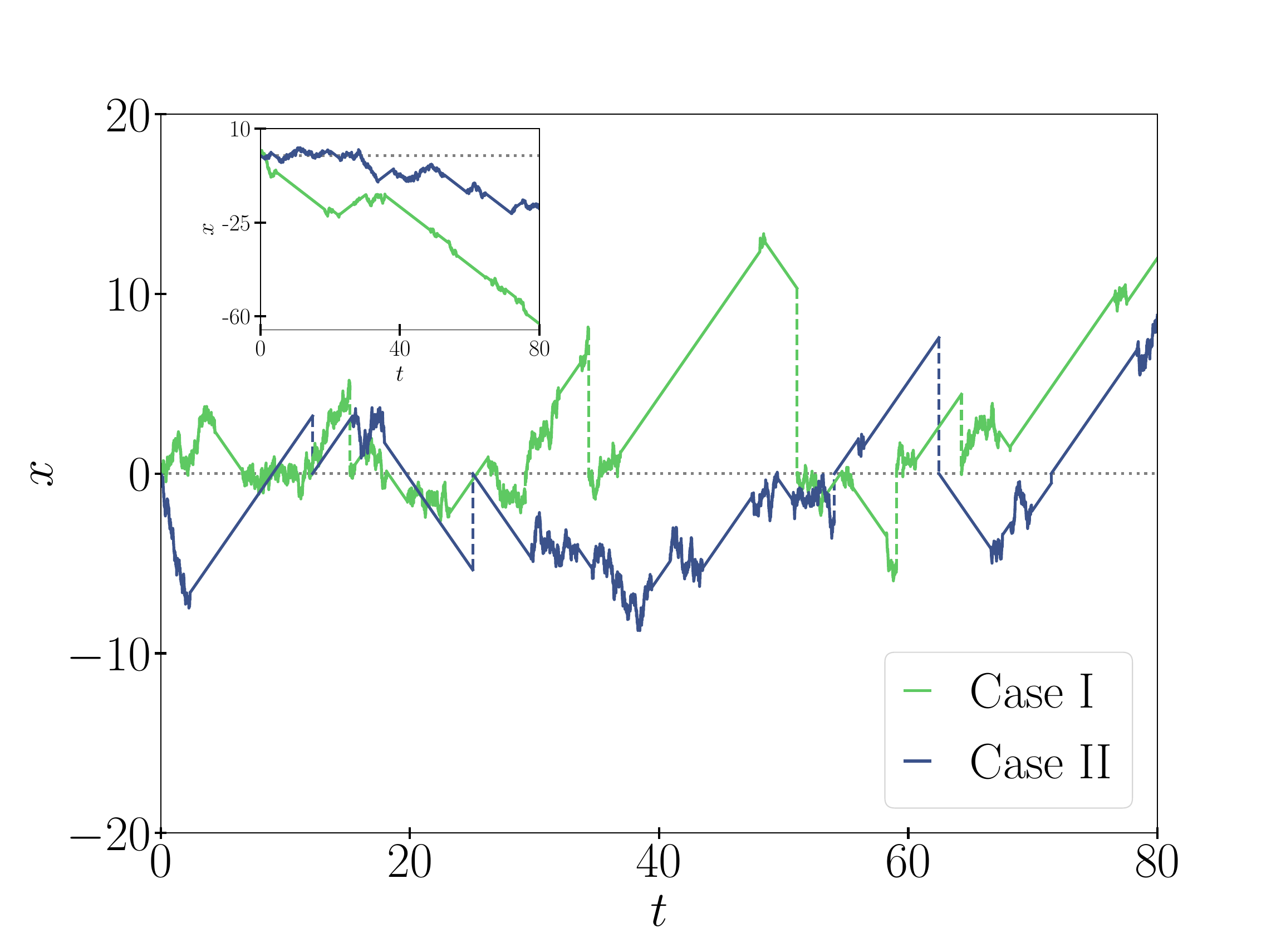} 
	\caption{Example of two trajectories the green one for case I and the blue one for case II. The dashed lines represent a reset to the initial position. The parameters used are $D=v=1$, $\tau_D = 3.8$, $\tau_v = 3.3$, $x_0=0$ and $r=0.1$. In the inset, other two trajectories for the case without reset, for case I (blue) and case II (green) respectively.}
	\label{fig:trajs}
\end{figure}

As we have shown, in unbounded domains the MFAT to a hidden target becomes infinity. We expect, and later show, that the presence of stochastic resetting to the initial position yields a finite MFAT to the target, as in numerous other works (see Refs. \cite{EM2011prl,EM2011} for example). The process with resetting is as follows: the walker initially starts at $x_0$ either diffusing (case I) or relocating (case II). After some random time  the walker is returned instantaneously to its initial position and the movement starts again. It is worth noting that a reset implies that the process is started all over again. This implies that if initially the walker started diffusing, after a reset the walker will be in the diffusing state, or vice versa if the walker is initially relocating. In Fig. \ref{fig:trajs} we present an example of two sampled trajectories of the IRW with stochastic resetting, where the dashed lines reveal that a reset has taken place (example trajectories without reset are included in the inset of the figure for better comparison).

\subsection{Stationary state}

 It is known \cite{Evans_2020}  that the propagator in the presence of resets, namely $\rho_r(x,t)$, satisfies the renewal equation 
\begin{align}
\rho_{r}(x,t)=\varphi^{*}(t)\rho(x,t)+\int_{0}^{t}\varphi(t')\rho_{r}(x,t-t')dt'
\label{re}
\end{align}
where $\rho(x,t)$ is the propagator in absence of resets, $\varphi(t)$ is the probability of a resetting at time $t$, and $\varphi^*(t)$ is the complementary probability of not having a reset up to time $t$, that is $\varphi^{*}(t)=\int_{t}^{\infty}\varphi(t')dt'$. Considering a Poissonian resetting process, i.e., exponentially distributed resets $\varphi(t) = \tau_r^{-1} e^{-t/\tau_r}$, the propagator in the presence of resets in the Fourier-Laplace space is related to the one without resets by means of:

\begin{align}
\hat{\rho}_{r}(k,s)=\frac{1+s\tau_r}{s\tau_r}\hat{\rho}\left(k,s+\frac{1}{\tau_r}\right).
\label{propr}
\end{align}

\begin{figure}[h!]
    \centering
    \includegraphics[width=\linewidth]{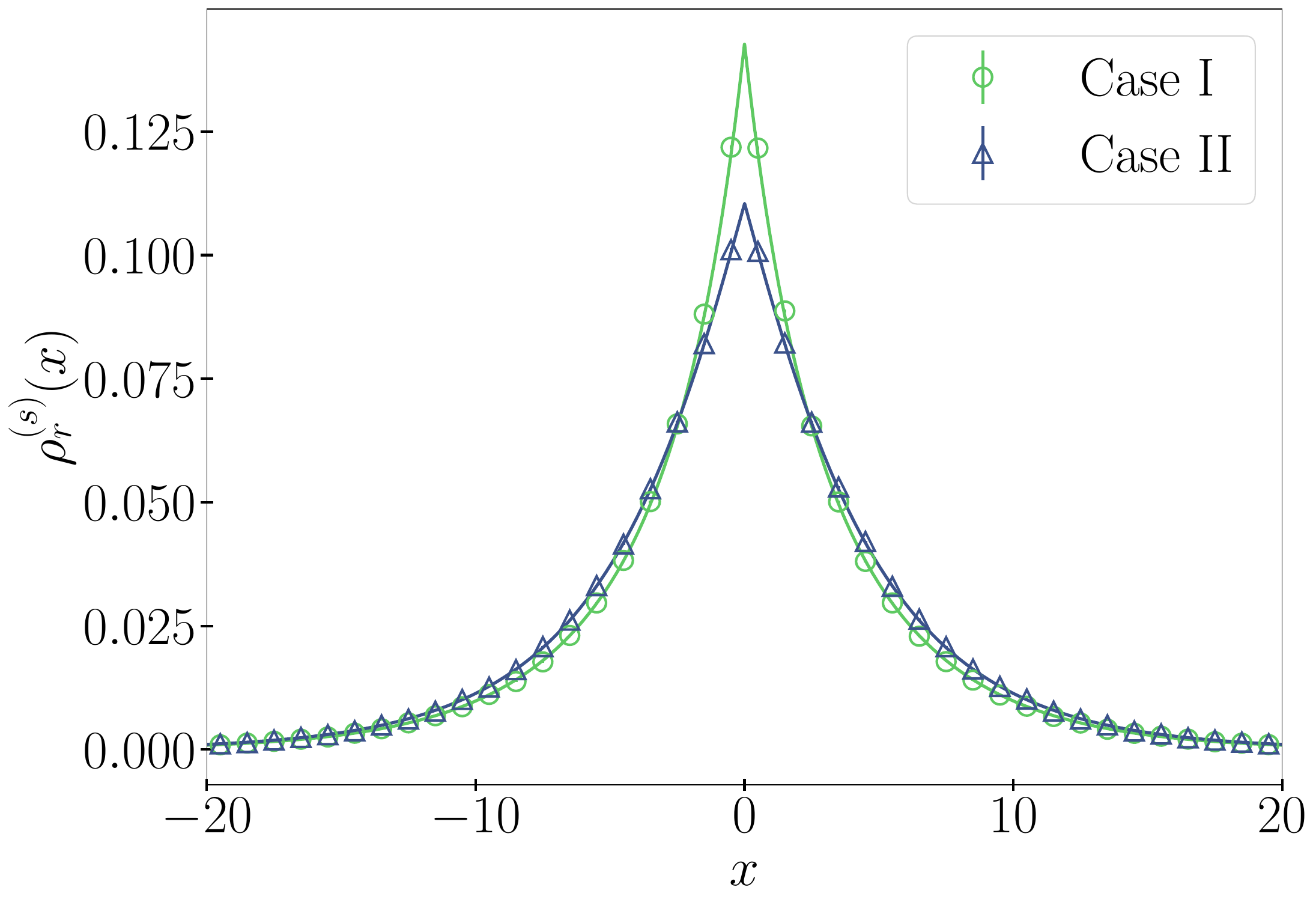}
    \caption{The propagator at the stationary state for both initial conditions, case I (green circles) and case II (blue triangles) in the system with resetting. In solid lines we present the analytic solution, Eq. \eqref{eq:steadydif} and Eq. \eqref{eq:steadyrel}, and in dots data obtained from numerical simulations. The parameters used are $D=v=1$, $\tau_D = 3.8$, $\tau_v = 3.3$, $x_0=0$ and $r=0.1$. }
    \label{fig:distrreset}
\end{figure}

It is known that the presence of resets induces a Non-Equilibrium Stationary State (NESS) \cite{EM2011prl,EM2011}. Hence, we compute the long time limit of Eq. \eqref{propr} in order to study the existence and properties of such NESS.

\begin{align}
    \bar{\rho}_{r}^{(s)}(k)=\lim_{s\to0}s\hat{\rho}_{r}(k,s)=r\hat{\rho}(k,s=r).
\end{align}
where $r=1/\tau_r$ is the constant reset rate and the superscript $(s)$ denotes that it is the stationary propagator. It is interesting to note that since the propagator depends on the initial state of the walker, Eqs. \eqref{wholeD} and \eqref{wholev}, the stationary state will also be different if the walker was initially either at the relocation or diffusive states. In case I,  

\begin{align}
 \rho_r^{(s),D}(x) & =  r\left[ \frac{1}{2\pi} \int_{-\infty}^{\infty}\hat{\rho}^D(k,s)e^{ikx}dk \right]_{s=r}  \nonumber \\
 & = r\left[ I(|x-x_0|, r) + \frac{\tau_v}{\tau_D}J(|x-x_0|, r)\right],
    \label{eq:steadydif}
\end{align}
where $I(x,s)$ is defined in Eq. \eqref{eq:I} and $J(x,s)$ in Eq. \eqref{eq:J(x0,s)}. We note that in the absence of relocations ($\tau_D\to\infty$) one recovers the result in Ref. \cite{EvMa11}. On the other hand, for case II the propagator at the stationary state is given by 

\begin{align}
    & \rho_r^{(s),v}(x) = r\left[ \frac{1}{2\pi} \int_{-\infty}^{\infty}\hat{\rho}^v(k,s)e^{ikx}dk \right]_{s=r}  \nonumber \\
    & = r\bigg[ J(|x-x_0|, r) + \frac{e^{-|x-x_0|\alpha}}{2v} \nonumber \\ 
    & + \frac{\tau_v(r\tau_v + 1)^2}{(2D\tau_D(v\tau_v)^4} \bigg( \frac{e^{-|x-x_0|\lambda_-}}{\lambda_-(\lambda_+^2-\lambda_-^2)(\alpha^2-\lambda_-^2)} \nonumber \\ 
    & - \frac{e^{-|x-x_0|\lambda_+}}{\lambda_+(\lambda_+^2-\lambda_-^2)(\alpha^2-\lambda_+^2)}  + \frac{e^{-|x-x_0|\alpha}}{\alpha(\lambda_+^2-\alpha^2)(\lambda_-^2-\alpha^2)}  \bigg) \bigg],
    \label{eq:steadyrel}
\end{align}
where $\alpha \equiv (r\tau_v+1)/(v\tau_v)$.  In Fig. \ref{fig:distrreset} we show the comparison between the NESS (computed through Eqs. \eqref{eq:steadydif} and \eqref{eq:steadyrel}) and numerical simulations, confirming the validity of our results. We can see that, as expected, the NESS is different depending on the initial condition of the walker. This is a result that we are going to recover for all other magnitudes computed in this work. This is due to the fact that resetting makes the walker return to the initial condition and so, its effect is not smoothed out at long times or long distances as is expected in IRW processes without resetting.

\subsection{Mean squared displacement}

\begin{figure}[h!]
    \centering
    \includegraphics[width=\linewidth]{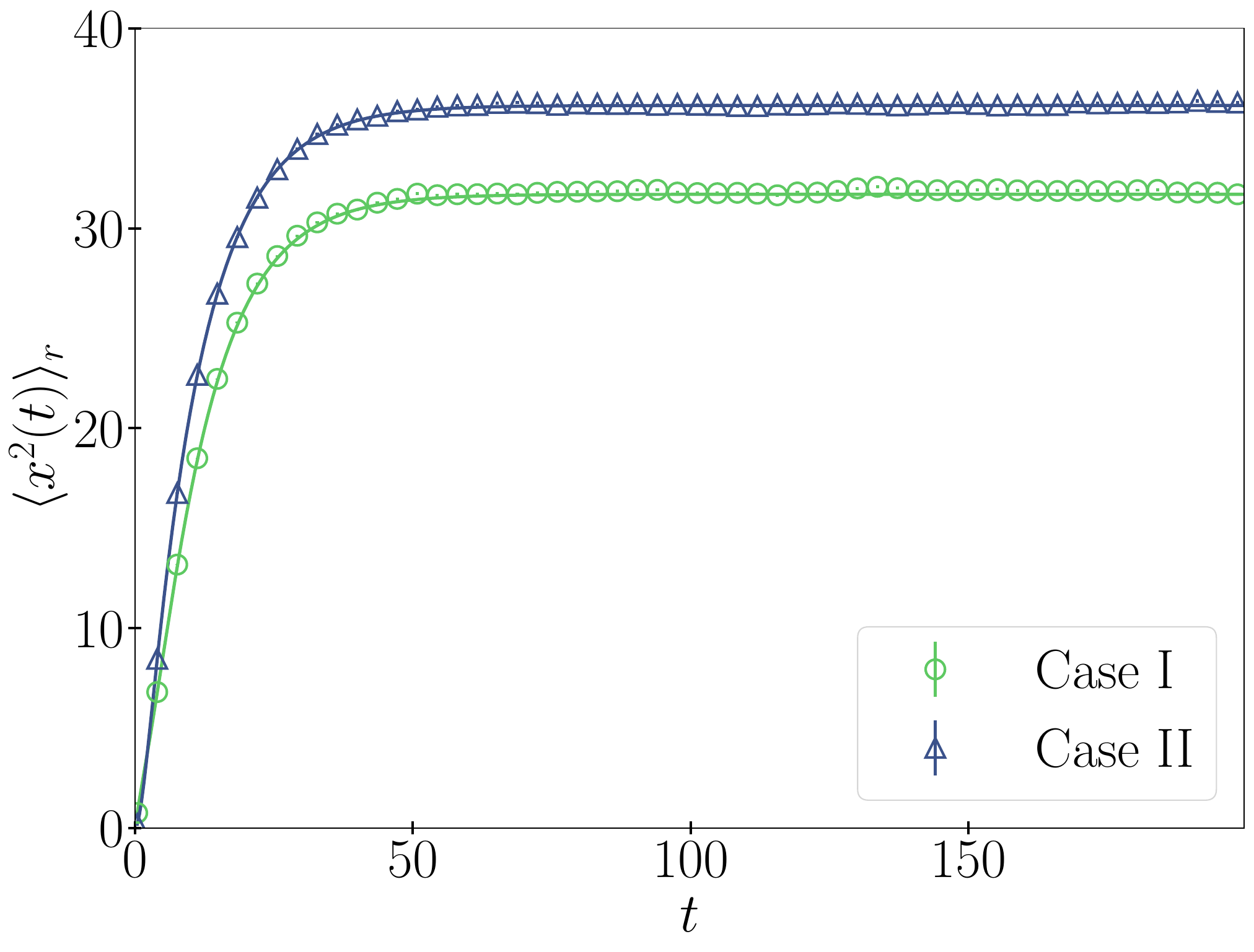}
    \caption{MSD in the presence of resetting for case I (green circles) and case II (blue triangles) for the system with resetting. In solid lines we present the analytic solution, Eq. \eqref{msdreset}, and in dots data obtained from numerical simulations. The parameters used being $D=v=1$, $\tau_D = 3.8$, $\tau_v = 3.3$, $x_0=0$ and $r=0.1$.}
    \label{fig:msdrest}
\end{figure}

Next, we compute the MSD when resetting is present in the model. Multiplying \eqref{re} by $x^2$ and integrating over $x$ one obtains
\begin{align}
\left\langle x^{2}(t)\right\rangle _{r}=\int_{0}^{t}Q(t-t')\varphi^{*}(t')\left\langle x^{2}(t')\right\rangle dt';
\label{msdre}
\end{align}
where $\left\langle x^{2}(t)\right\rangle $ is the MSD in absence of resetting and $Q(t)$ is the resetting rate function whose Laplace transform is
$$
Q(s)=\frac{1}{1-\varphi(s)}=\frac{1+s\tau_{r}}{s\tau_{r}}.
$$
Inserting  \eqref{msdsr} into \eqref{msdre} we finally obtain that the MSD in the presence of resetting is given by

\begin{align}
        \left\langle x^{2}(t)\right\rangle_{r}&= 2\frac{D+\frac{\tau_v^2v^{2}}{\tau_D}}{(1+\frac{\tau_v}{\tau_D})}\tau_{r}\left(1-e^{-t/\tau_{r}}\right) \nonumber\\     &+A_{1}+\frac{A_{2}}{1+\frac{\tau_{r}}{\tau_v}}\left(1+\frac{\tau_{r}}{\tau_v} e^{-t\frac{(\tau_r + \tau_v)}{\tau_r\tau_v}}\right)\nonumber\\     &+\frac{A_{3}\left[1+\frac{\tau_{r}}{\tau_v}\left(1+\frac{\tau_v}{\tau_D}\right)e^{-t\frac{(\tau_r + \tau_v + \tau_D)}{\tau_D\tau_r\tau_v}}\right]}{1+\frac{\tau_{r}}{\tau_v}\left(1+\frac{\tau_v}{\tau_D}\right)}
\label{msdreset}
\end{align}
where again the constants $A_1$, $A_2$ and $A_3$ are as in \eqref{msdA1d} - \eqref{msdA3d} for case I and \eqref{msdA1r} - \eqref{msdA3r} for case II. From Eq. \eqref{msdreset} it can be easily checked that 
\begin{eqnarray*}
\left\langle x^{2}(t)\right\rangle_{r}&=& 2\frac{D+\frac{\tau_v^2v^{2}}{\tau_D}}{(1+\frac{\tau_v}{\tau_D})}\tau_{r} +A_{1}+\frac{A_{2}}{1+\frac{\tau_{r}}{\tau_v}}\\
&+&\frac{A_{3}}{1+\frac{\tau_{r}}{\tau_v}\left(1+\frac{\tau_v}{\tau_D}\right)},\; \textrm{as}\; t\to\infty.
\end{eqnarray*}
so $\left\langle x^{2}(t)\right\rangle_{r}$ tends to a constant value as time goes to infinity, which is expected since it is the second moment of the NESS. In Fig. \ref{fig:msdrest} we compare our analytical result \eqref{msdreset} with numerical simulations and see that the analytical expressions are in good agreement with the numerical data.

\subsection{FAT under resetting}

 For the case without resetting, we have seen that the MFAT goes to infinity for a target located in an unbounded space. But as we have mentioned above many systems can optimize the MFAT in the presence of stochastic resetting. Following this reasoning we proceed to compute the FATD for the IRW under resetting. We do so in Laplace space in terms of the FATD of the system without resetting. Following \cite{Evans_2020}, let $G_0(x_0,t)$ be the survival probability for the system without resetting and $G_r(x_0,t)$ the one for the system with resetting; both magnitudes are related via a last renewal equation

\begin{align}
    G_r(x_0,t) & =   e^{-rt} G_0(x_0,t)  \nonumber\\  
    & +  r\int_0^t  e^{-r\tau}G_0(x_0,\tau)G_r(x_0,t-\tau)d\tau,
    \label{eq:rensurviv}
\end{align}

\begin{figure}[h!]
	\centering
	\includegraphics[width=\linewidth]{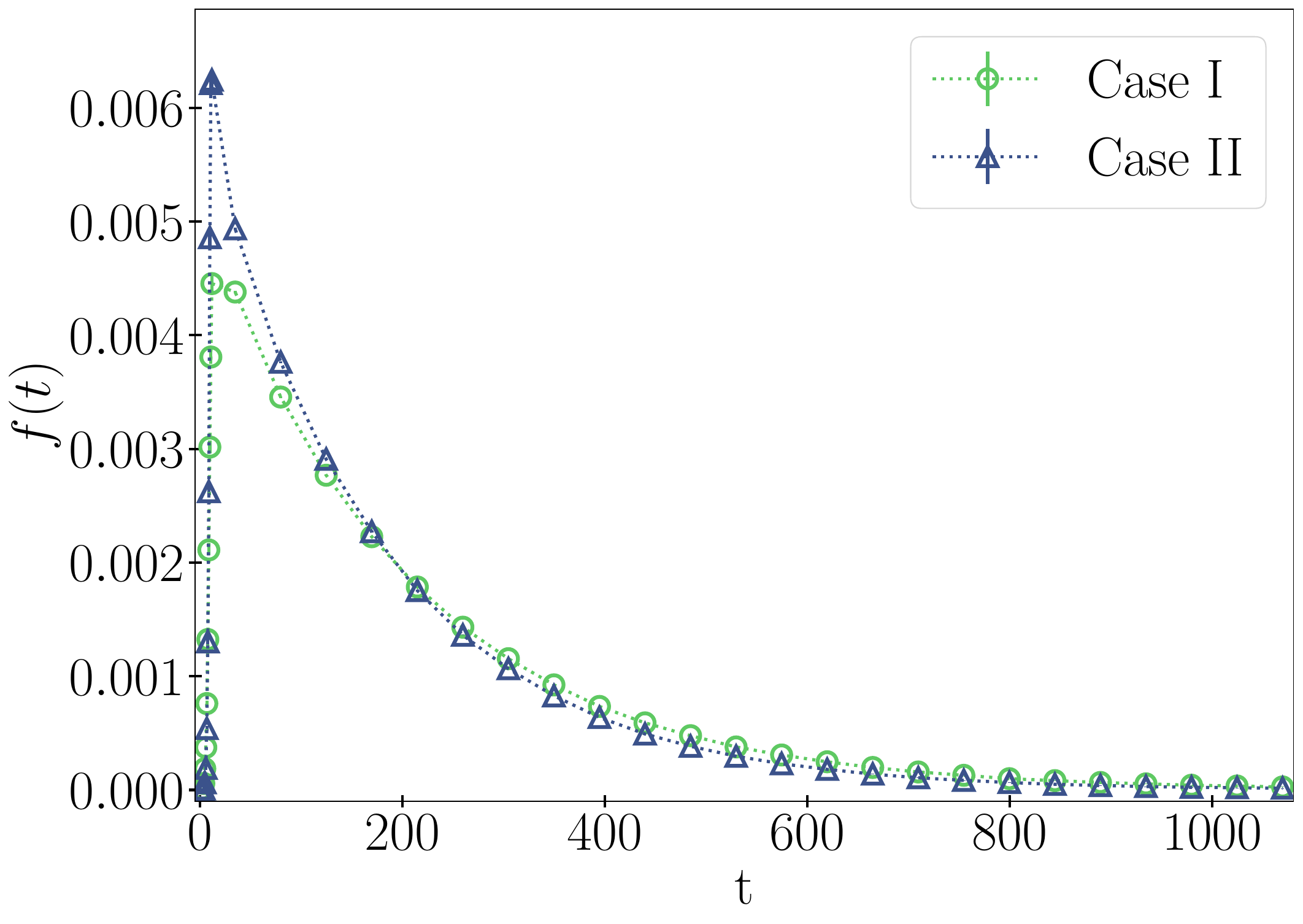} 
	\caption{FATD in the presence of resetting for both case I (green circles) and case II (blue triangles) obtained from numerical simulations. The dotted lines are added as a visual guide. The parameters used being $D=v=1$, $\tau_D = 3.8$, $\tau_v = 3.3$, $x_0=10$ and $r=0.1$.}
	\label{fig:FATDr0}
\end{figure}

Eq. \eqref{eq:rensurviv} in Laplace space becomes
	
\begin{align}
	\tilde{G}_r(x_0,s) = \frac{\tilde{G}_0(x_0,s+r)}{1-r\tilde{G}_0(x_0,s+r)}.
	\label{eq:qr-q0}
\end{align}
In order to find the FATD, $f(t)$, we note that it is related to the survival probability via $f(t) = - \frac{\partial G(x_0,t)}{\partial t}$, so, from Eq. \eqref{eq:qr-q0} the FATD with resetting in Laplace space reads

\begin{align}
	\tilde{f}_r(s) = \frac{(s+r)\tilde{f}_0(s+r)}{s+r\tilde{f}_0(s+r)},
    \label{eq:FATD_r}
\end{align}
where $\tilde{f}_0(s)$ is the first passage probability density without resetting. In principle \eqref{eq:FATD_r} can be transformed back to real space, via the inverse Laplace transform. In Fig. \ref{fig:FATDr0} we present a particular case for the FATD for case I and II in real space from simulation data. We have checked that the FATD decays exponentially, though there is not a simple relation between the decaying constant and the model's parameters. 

We can further compute the MFAT $\left\langle T (x_0) \right\rangle _r$  from the first passage probability density

\begin{align}
	     \left\langle T (x_0) \right\rangle _r &=  \int_{0}^{\infty} t f_r(x_0,t)dt=\tilde{G}_r(x_0,s=0)  \nonumber \\ 
	    &= \frac{\tilde{G}_0(x_0,r)}{1-r\tilde{G}_0(x_0,r)} = \frac{1 -\tilde{f}_0(s=r)}{r\tilde{f}_0(s=r)}.
	\label{eq:MFAT-step1}
\end{align}

Combining Eqs. \eqref{eq:FATD_d} and \eqref{eq:MFAT-step1} the expression of the MFAT for the case I is given by

\begin{align}
        \left\langle T^{D}(x_{0}) \right\rangle _r =   \frac{1}{r}\left[\frac{\phi_{-}-\phi_{+}}{\phi_{-}e^{-|x_{0}|\lambda_{-}}-\phi_{+}e^{-|x_{0}|\lambda_{+}}}-1\right]
    \label{eq:MFAT1}
\end{align}
where $\phi_{+}$, $\phi_{-}$, $\lambda_{+}$ and $\lambda_{-}$ are as in Eqs. \eqref{eq:lambda} and \eqref{fis} but setting $s=r$.
It can be seen that if the searcher does not relocate ($\tau_D\rightarrow\infty$), i.e., only diffuses then $\lambda_+\simeq \sqrt{r/D}$ and $\phi_{-}\simeq 0$. Therefore, \eqref{eq:MFAT1} reduces to the well-known case of Brownian walkers with stochastic resetting \cite{EvMa11}. Analogously, for case II the MFAT reads

\begin{align}
       \langle T^{v}(x_{0})\rangle_r=\frac{1}{r}\bigg[\frac{(v\tau_{v})^{2}(\phi_{-}-\phi_{+})/(r\tau_{v}+1)}{\lambda_{-}^{-1}e^{-|x_{0}|\lambda_{-}}-\lambda_{+}^{-1}e^{-|x_{0}|\lambda_{+}}}-1\bigg].
   \label{eq:MFAT2}
\end{align}
We note that this result is different from that obtained in Eq. \eqref{eq:MFAT1} which reveals the explicit dependence of the MFAT on the initial condition, i.e., if the walker is initially either relocating or diffusing. In this case, in the limit of no relocation, $\tau_D \to \infty$,  the MFAT goes to

\begin{equation} 
\left\langle T^{v}(x_{0})\right\rangle \simeq \frac{1}{r}\left[ \frac{\lambda_+^2 - \lambda_-^2}{\lambda_+e^{-|x_0|\lambda_-} + \lambda_-e^{-|x_0|\lambda_+}} - 1 \right].
\end{equation}

This shows that the MFAT is dependent on the initial condition also in this limit. It can easily be understood if noticing that $\tau_D \to \infty$ means that the searcher only diffuses, so for case I the walker does not relocate. It is not the case for case II, where the walkers relocate at rate $\tau_v^{-1}$ before switching to diffusion state. In the opposite limit, when $D=0$ the searcher only relocates i.e., it never detects the target so that the MFAT \eqref{eq:MFAT1} and \eqref{eq:MFAT2} diverge. Effectively, if $D=0$ then $\lambda_\pm \rightarrow\infty$ and $\left\langle T(x_{0})\right\rangle_r\rightarrow \infty$. One may notice that while $\left\langle T^{D}(x_{0})\right\rangle_r \to 0$ as $x_0 \to 0$, $\langle T^{v}(x_{0})\rangle_r$ tends to a constant value given by

\begin{equation} 
 \langle T^{v}(x_{0})\rangle_r\simeq\frac{1}{r}\bigg[\frac{(v\tau_{v})^{2}}{r\tau_{v}+1}\frac{(\phi_{-}-\phi_{+})\lambda_{+}\lambda_{-}}{\lambda_{+}-\lambda_{-}}-1\bigg]
\end{equation}
in the same limit.
This difference arises because for case II the walkers do a relocation first (during which target detection is not possible), so in general this is detrimental in order to find the target  for $x_0\to0$. More interesting is the case for $x_0\to\infty$, which satisfies

\begin{equation}  
\frac{\langle T^{D} (x_0)\rangle_r }{\langle T^{v} (x_0)\rangle_r} \simeq \frac{r\tau_v + 1}{(v\tau_v)^2}\frac{1}{\phi_-\lambda_-} .
\label{eq:quocient}
\end{equation}

 In general, $\langle T^{D}(x_{0})\rangle_{r}\neq \langle T^{v}(x_{0})\rangle_{r}$ showing that the effect of the initial condition is not smoothed out by a long time evolution. As we have mentioned this behavior can be understood because, in the presence of resetting, the walker is consistently returned to its initial condition during the time evolution.

\begin{figure}[h!]
	\centering
	\includegraphics[width=\linewidth]{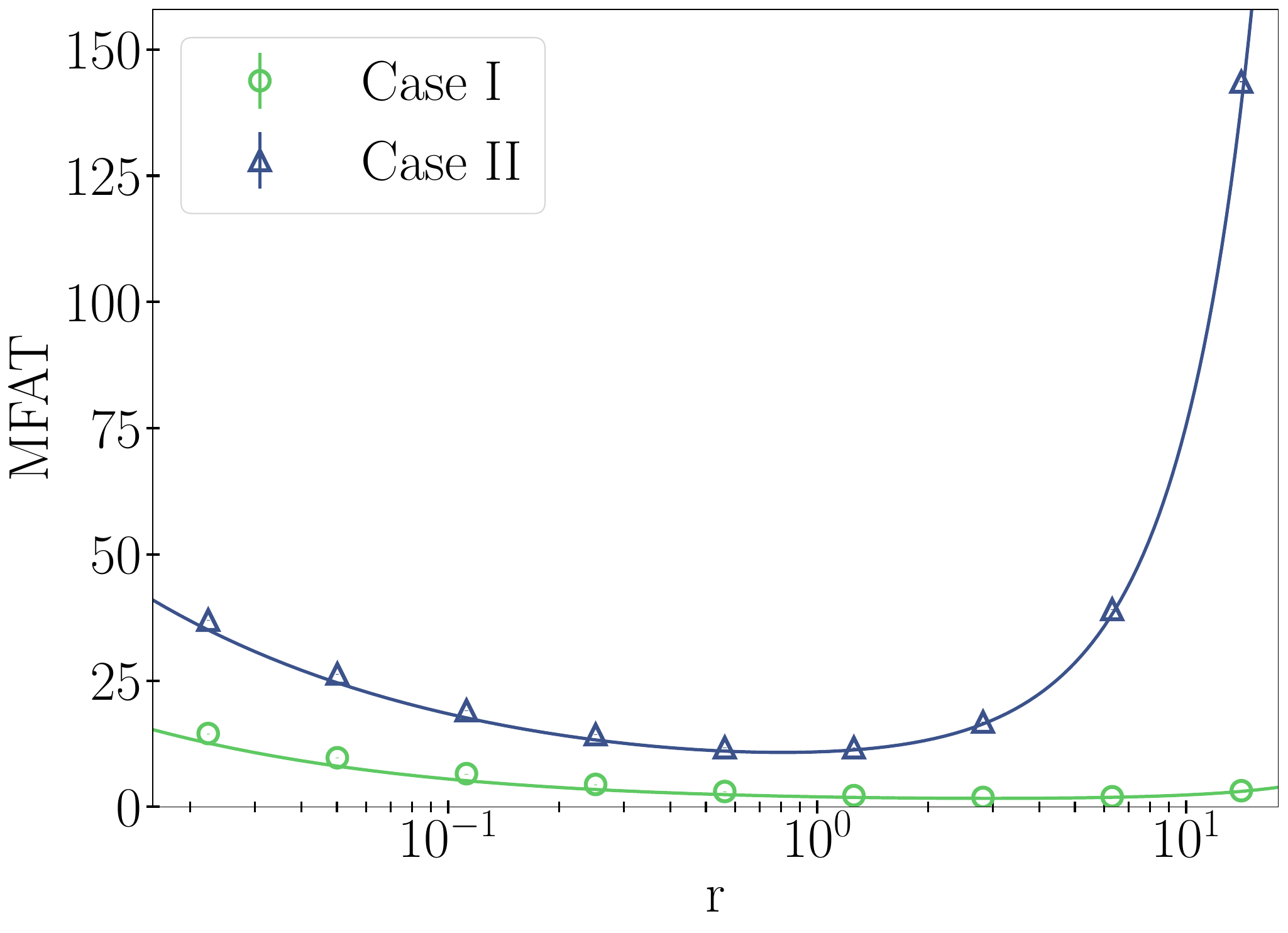} 
	\caption{MFAT in the presence of resetting for both case I (green circles) and case II (blue triangles). In solid lines we show Eq. \eqref{eq:MFAT1} and  Eq. \eqref{eq:MFAT2} and in dots data obtained from numerical simulations. The parameters used being $D=v=1$, $\tau_D = 3.8$, $\tau_v = 3.3$, $x_0=10$ and $r=0.1$.}
	\label{fig:FATDr}
\end{figure}

\subsection{Existence of an optimal reset rate}

\begin{figure}
	\includegraphics[width = \columnwidth]{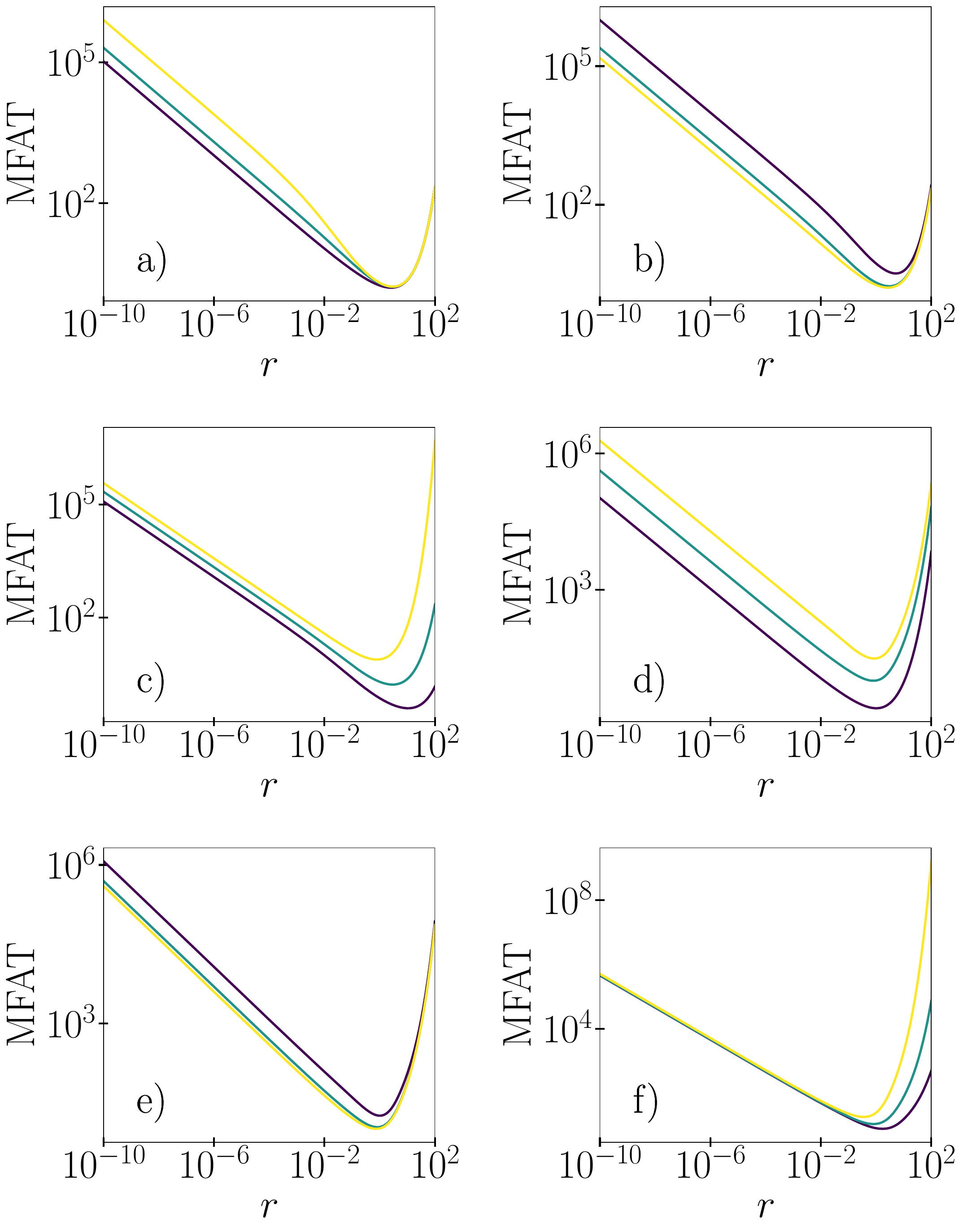}[h!]
	\caption{MFAT vs $r$ in the presence of resetting for different parameter values. In panels a), b) and c) case I is represented, the MFAT computed with Eq. \eqref{eq:MFAT1}; Panels d), e) and f) are for case II, its MFAT computed with Eq. \eqref{eq:MFAT2}. The parameters used are $D=v=1$, $\tau_D = 3.8$, $\tau_v = 3.3$ and $x_0=1$ unless stated the contrary. In panels a) and d) the purple curve is for $\tau_v = 0.3$, the blue one for $\tau_v = 3$ and the yellow one for $\tau_v = 10$. For panels b) and e) the purple curve is for $\tau_v = 0.3$, the blue one for $\tau_v = 3$ and the yellow one for $\tau_v = 10$. In panels c) and f) the purple curve is for $x_0 = 0.5$, the blue one for $x_0 = 1$ and the yellow one for $x_0 = 2$. All axes are on a logarithmic scale.}
	\label{fig:MFATr}
\end{figure}

In Fig. \ref{fig:FATDr} we compare Eq. \eqref{eq:MFAT1} and Eq. \eqref{eq:MFAT2} to numerical simulations, so we can observe the emergence of an optimal value of $r$ minimizing the MFAT of the IRW, as happens for the case of Brownian walkers too \cite{EvMa11}. In the absence of resetting the walker takes an infinite time to reach the target due to the non-recurrent nature of Brownian particles. On the other hand, for large reset rates the walker's motion is localized near the reset point, also resulting in a very large MFAT. 
To explore analytically the existence of this optimal rate, we analyze  Eq. \eqref{eq:MFAT1} both for small and large $r$. Let us first expand the MFAT for large $r$. From Eqs. \eqref{eq:B} and \eqref{eq:C} we have $B(r)\simeq r^2/v^2$ and $C(r)\simeq r^3/Dv^2$ so that
\begin{align}
\lambda_{+}\simeq r/v,\;\lambda_{-}\simeq\sqrt{r/D}\quad\text{as}\quad r\rightarrow\infty
\label{ls}
\end{align}
Finally, 
\begin{align}
\left\langle T^D(x_{0})\right\rangle_r \simeq \frac{1}{r}e^{|x_0|\sqrt{r/D}}\quad\text{as}\quad r\rightarrow\infty
\end{align}
so that the MFAT tends to infinity as $r$ does. This is the same behaviour as in the absence of resetting. To prove that the optimal reset rate exists we just need to verify if the MFAT decreases with $r$ for small values of $r$, this is, if the condition 
\begin{align}
\lim_{r\rightarrow0}\left(\frac{\partial\left\langle T^D(x_{0})\right\rangle }{\partial r}\right)<0
\label{cond}
\end{align}
is met. Expanding $\lambda_\pm$ for small $r$ we find 
\begin{align}
\lambda_+\simeq \lambda_0+ O(r),\quad \lambda_-\simeq \lambda_1\sqrt{r}+O(r),
\label{ls2}
\end{align}
where
$$
\lambda_{0}=\sqrt{\frac{1}{D\tau_{D}}+\frac{1}{(v\tau_{v})^{2}}},\quad \lambda_1=\frac{\tau_{D}+\tau_{v}}{(v\tau_{v})^{2}+D\tau_{D}}.
$$
After some calculations, the dominant term in the expansion of the MFAT for small $r$ reads
\begin{align}
\left\langle T^D(x_{0})\right\rangle \simeq \frac{T_0}{\sqrt{r}}\quad\text{as}\quad r\rightarrow 0,
\label{mtap}
\end{align}
where
$$
T_{0}=\lambda_{1}(v\tau_{v})^{2}\left[\left(e^{-|x_{0}|\lambda_{0}}-1\right)\left(\frac{1}{\lambda_{0}(v\tau_{v})^{2}}-\lambda_{0}\right)+\frac{|x_{0}|}{(v\tau_{v})^{2}}\right].
$$
Finally, the condition \eqref{cond} applied to \eqref{mtap} is equivalent to $T_0>0$, or simply
$$
\frac{|x_{0}|\lambda_{0}}{1-e^{-|x_{0}|\lambda_{0}}}>-\frac{(v\tau_{v})^{2}}{D\tau_{D}},
$$
which is always fulfilled. This confirms the existence of the optimal reset rate. 

One can also prove that \eqref{eq:MFAT2} has an optimal reset rate following an analogous procedure. In the limit of large reset rate $r$, using \eqref{ls} we find that
\begin{align}
\left\langle T^v(x_{0})\right\rangle \simeq \tau_r e^{|x_0|\sqrt{r/D}}\quad\text{as}\quad r\rightarrow\infty.
\end{align}
Analogously, using \eqref{ls2} we find that when $r\rightarrow 0$ the MFAT is given by \eqref{eq:MFAT2} and behaves as in \eqref{mtap} but with 
$$
T_{0}=\frac{\lambda_{1}}{\lambda_{0}}\left[\frac{(v\tau_{v})^{2}}{D\tau_{D}}+\lambda_{0}|x_{0}|+e^{-|x_{0}|\lambda_{0}}\right].
$$

Since $T_0$ is positive, the optimal reset rate also exists in this case. In Fig. \ref{fig:MFATr} we present some illustrative cases of the dependence of the MFAT with $r$ computed from Eq. \eqref{eq:MFAT1} and Eq. \eqref{eq:MFAT2} for case I and case II, respectively, to check the existence of the optimal value of $r$. Though we are not able to find a compact relationship between the model parameters and the minimal MFAT, in the next Section we identify numerically the existence of different regime behaviours of the MFAT.

\subsection{Phase space parameter exploration}\label{sec:parameter exploration}

\begin{figure}[h!]
    \includegraphics[width=\columnwidth]{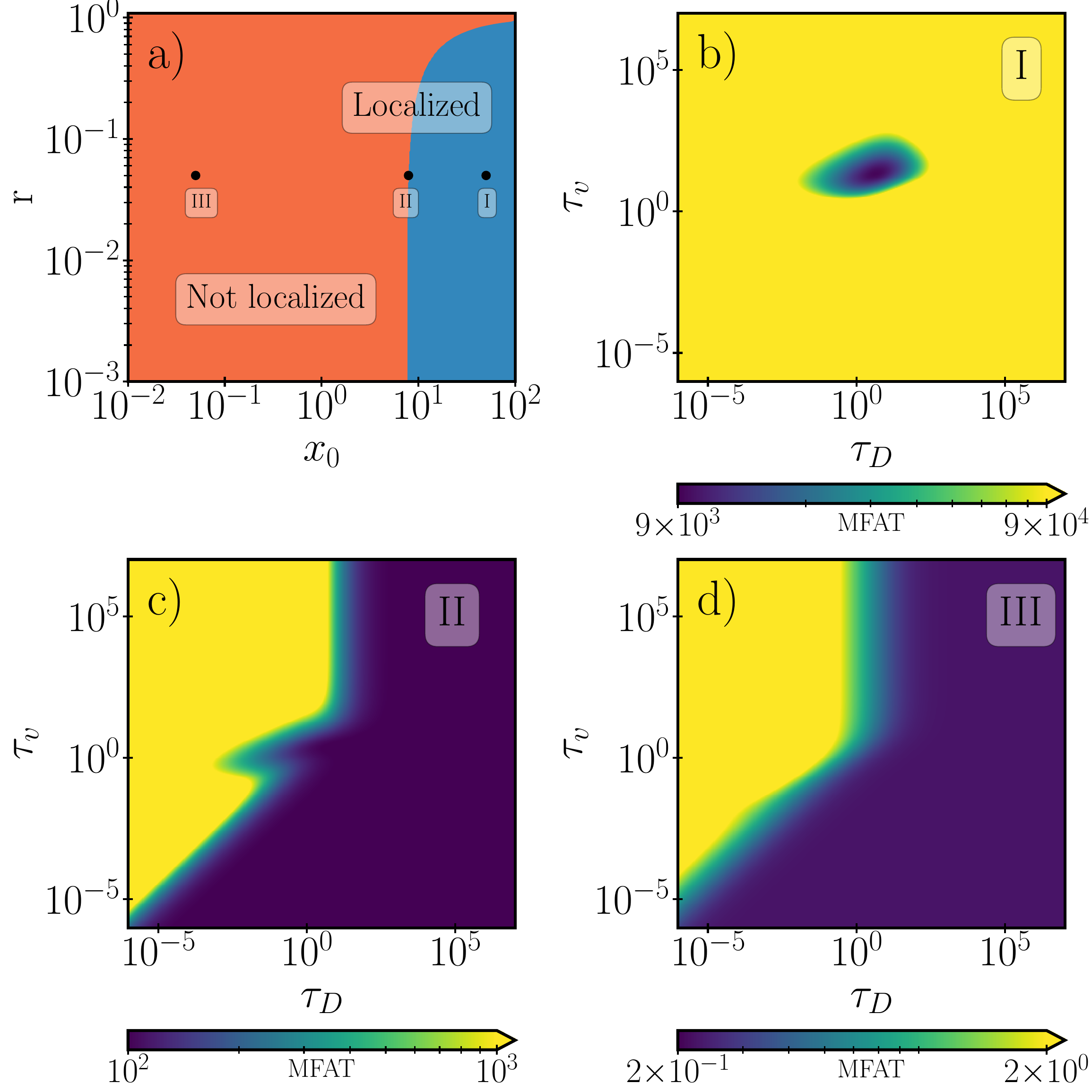}
    \caption{Panel a) Phase diagram of the behavior of MFAT in $(x_0,r)$ space for the diffusive initial condition (case I). Panels b), c), and d) are representative examples of the MFAT in each phase and the interphase in $(\tau_D, \tau_v)$ space for a given $x_0$ and $r$. The Roman number shows their corresponding position in the phase diagram. In panel b) $x_0=50$, $r=0.05$; In panel  c) $x_0=7.9$, $r=0.05$; And in panel d) $x_0=0.05$, $r=0.05$. For all plots $D=1$ and $v=1$.}
    \label{fig:phases_diff}
\end{figure}

\begin{figure}[h!]
    \includegraphics[width=\columnwidth]{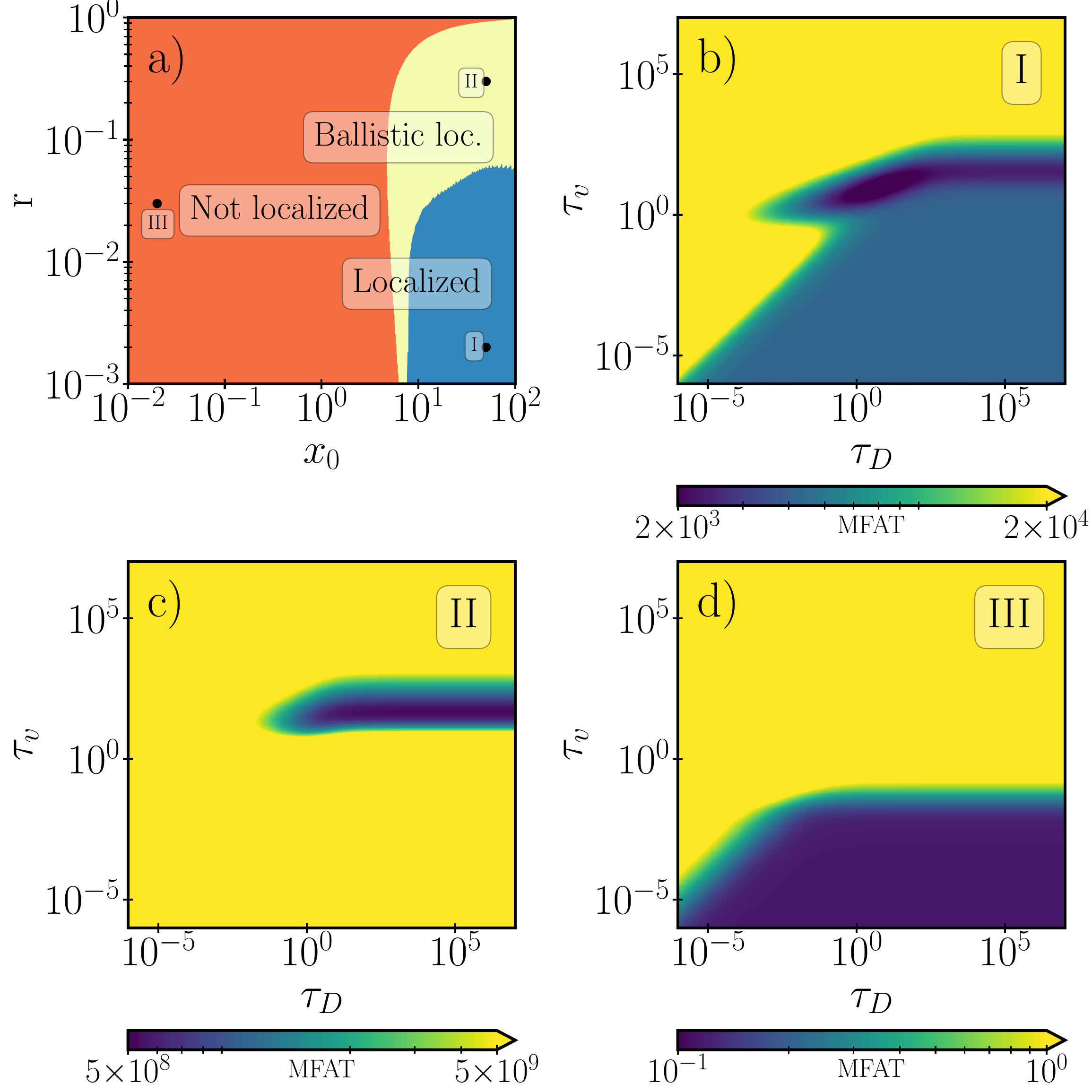}
    \caption{Panel a) Phase diagram of the behavior of MFAT in $(x_0,r)$ space for the relocation initial condition (case I). Panels b), c), and d) are representative examples of the MFAT in each phase in $(\tau_D, \tau_v)$ space for a given $x_0$ and $r$. The Roman number shows their corresponding position in the phase diagram. In panel b) $x_0=50$, $r=0.002$; In panel  c) $x_0=50$, $r=0.3$; And in panel d) $x_0=0.02$, $r=0.03$. For all plots $D=1$ and $v=1$.}
    \label{fig:phases_vel}
\end{figure}

We have numerically studied the MFAT behavior for distinct parameter values $\tau_D$, $\tau_v$, $x_0$, and $r$. We have kept $v=1$ and $D=1$ (note that one may change $x$ and $t$ to nondimensional variables $x \to x \frac{v}{D}$ and $t \to t \frac{v^2}{D}$). This allows us to focus the study on the most interesting regime in which ballistic and diffusive movements are in direct competition.

We have centered our study on finding the combination of parameter values that globally minimize the MFAT. The analysis was conducted in the following manner: for a set of $(\tau_D,\tau_v)$ in the range from $10^{-4}$ to $10^7$ we used Eq. \eqref{eq:MFAT1} and Eq. \eqref{eq:MFAT2} to compute the MFAT for different combinations within the ranges $r \in [10^{-3},10^0]$ and $x_0 \in [10^{-2},10^2]$.

For the diffusive initial condition, two distinct phases can be observed (distinguished by the colors red and cyan in Fig. \ref{fig:phases_diff}a). For the phase in cyan, which we name the \textit{localized phase}, the MFAT presents a well-localized minimum in the $(\tau_D,\tau_v)$ phase space. This can be better identified in Fig. \ref{fig:phases_diff}b, where we directly show for a specific choice of $(x_0,r)$ values that the minimum value lies in a confined region of the space ($\tau_D, \tau_v$). On the other hand, for $x_0$ small we obtain the red (\textit{non-localized}) phase, for which the MFAT does not have a well-defined minimum but rather there exists a plateau region of $(\tau_D, \tau_v)$ values for which the MFAT becomes minimum. In particular, any combination of $(\tau_D, \tau_v)$ out of the region $\tau_v \gg \tau_D$ will suffice to optimize the MFAT, as can be seen for the case depicted in Fig.\ref{fig:phases_diff}d. For the sake of completeness, in panel \ref{fig:phases_diff}c we show the behavior of the MFAT for the interface between both (localized and non-localized) phases, where we observe the transition from one behavior to the other.

In Fig. \ref{fig:phases_vel} we present the same analysis for the relocation initial condition, case II. In this case, we observe three different phases. Apart from the localized and non-localized phases above, we observe an intermediate situation (which we name \textit{ballistic localization}, in light yellow in Fig. \ref{fig:phases_vel}a). For this region, the minimum of the MFAT is confined to a fine line of $(\tau_D,\tau_v)$ values where $\tau_v$ is (regardless of the value of $\tau_D$) at least of the order of the timescale $x_0/v$, which corresponds to the case where the target can be reached on average in a single relocation. Similarly to case I, in panels b-d of Fig. \ref{fig:phases_vel} we show the behavior of the MFAT for the three different phases. Remarkably, the behavior for the non-localized phase is slightly different in that now the region where the minimum MFAT occurs does not include the region where $\tau_v$ is large. This is because for $x_0$ small a very large value of $\tau_v$ now implies that the walker will depart on average much from the target position before switching to diffusive motion when the target can be detected.

\begin{figure}[htbp]
    \includegraphics[width=\columnwidth]{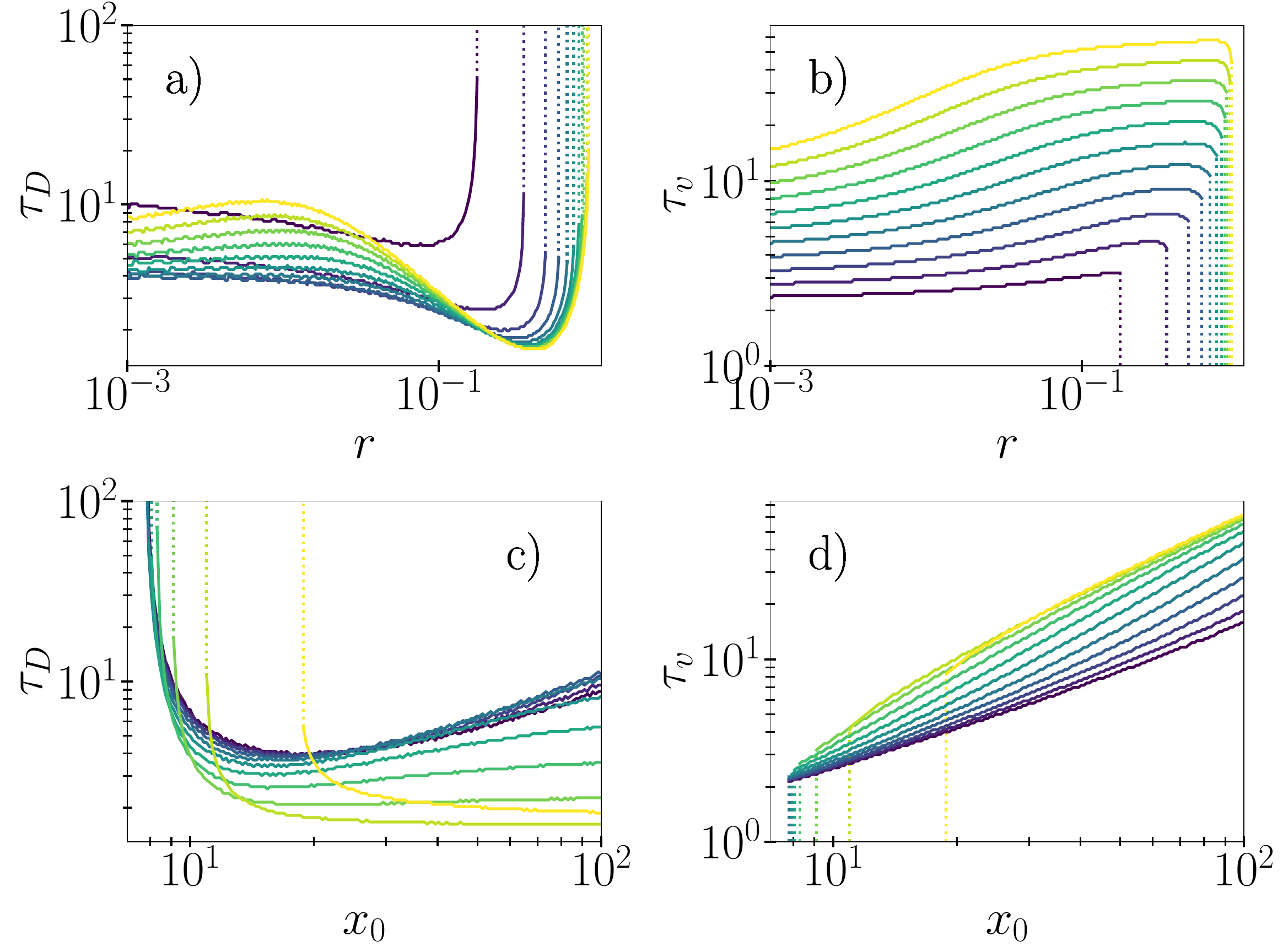}
	\caption{Plots of $\tau_D$ and $\tau_v$ that minimize the MFAT for the localized phase of case I. The MFAT has been computed with Eq. \eqref{eq:MFAT1}. Panels a) and b) represent $\tau_D$ and $\tau_v$ respectively as a function of $r$, the different curves for different values of $x_0 = 9.1,$$11.4,$$14.4,$$18.2,$$23.0,$$ 29.0,$$36.6,$$46.1,$$58.2,$$73.4,$ and $ 92.5$ from purple to yellow. Panels c) and d) represent $\tau_D$ and $\tau_v$ respectively as a function of $x_0$, the different curves being for different values of $r$, again, the purple is for smaller values of $r$ and yellow for larger ones. The $r$ represented are $r = 0.001,$$ 0.0018,$$ 0.0035,$$ 0.0068,$$ 0.012,$$ 0.024,$$ 0.046,$$ 0.088,$$ 0.16,$$ 0.31,$ and $ 0.6$. The dotted line on all panels signals the transition to the other phase. All axes are on a logarithmic scale.}
    \label{fig:minimum_dif}
\end{figure}

\begin{figure}[htbp]
    \includegraphics[width=\columnwidth]{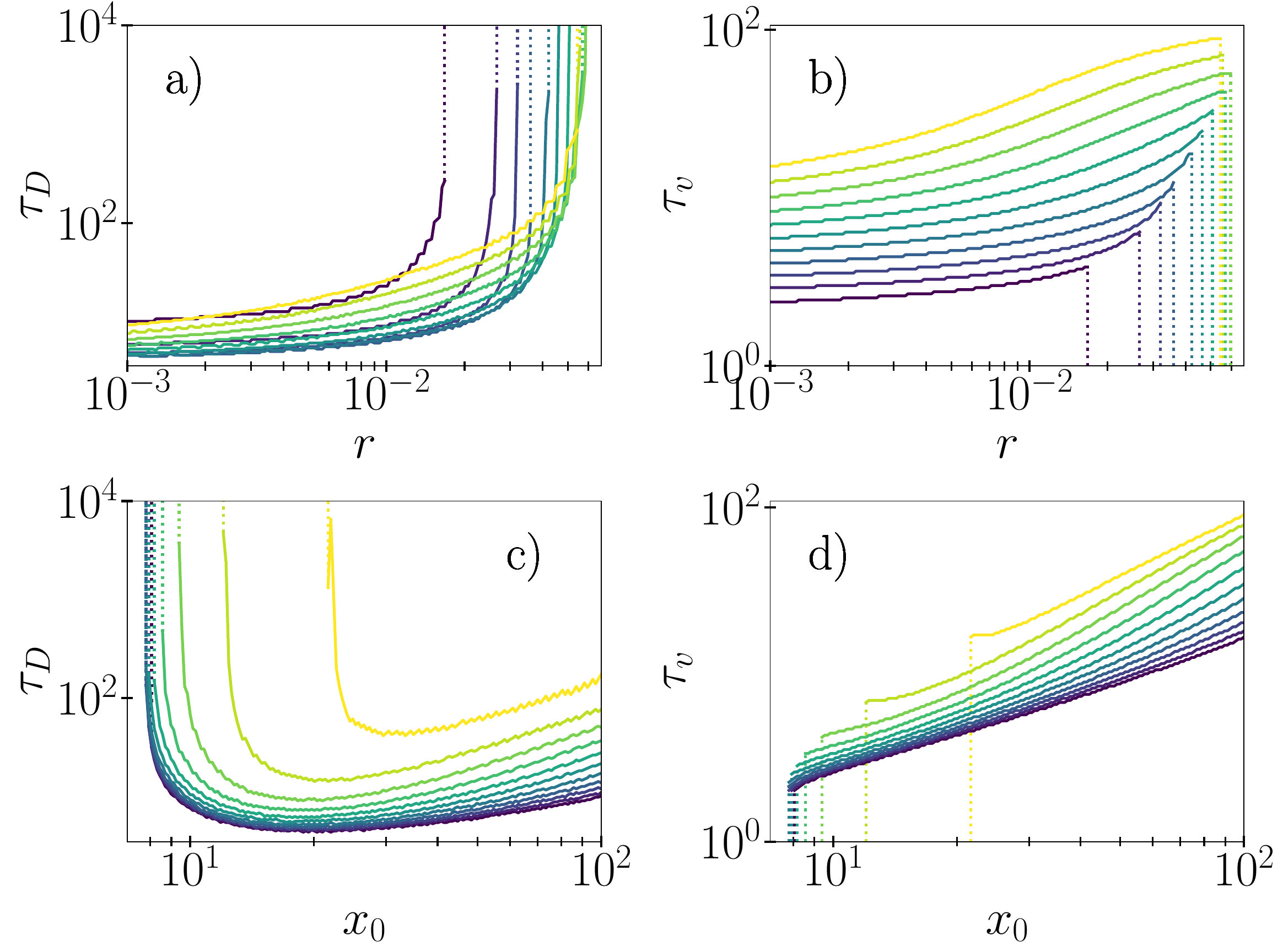}
	\caption{Plots of $\tau_D$ and $\tau_v$ that minimize the MFAT for the localized phase of case II. The MFAT has been computed with Eq. \eqref{eq:MFAT2}. Panels a) and b) represent $\tau_D$ and $\tau_v$ respectively as a function of $r$, the different curves for different values of $x_0 = 9.1,$$11.4,$$14.4,$$18.2,$$23.0,$ $29.0,$$36.6,$$46.1,$$58.2,$$73.4,$ and $ 92.5$ from purple to yellow. Panels c) and d) represent $\tau_D$ and $\tau_v$ respectively as a function of $x_0$, the different curves being for different values of $r$, again, the purple is for smaller values of $r$ and yellow for larger ones. The $r$ represented are $0.001,$$ 0.0014,$$ 0.002,$$ 0.003,$ $ 0.0043,$$ 0.0063,$$ 0.0091,$$ 0.013,$$0.019,$$ 0.027,$ and $0.04$. The dotted line on all panels signals the transition to the other phase. All axes are on a logarithmic scale.}
    \label{fig:minimum_vel}
\end{figure}

In parallel, we have studied how the values $\tau_D,\tau_v$ which minimize the MFAT depend on the model parameters $x_0,r$ in the localized phase for both initial conditions. The corresponding numerical results are presented in Figs. \ref{fig:minimum_dif} and \ref{fig:minimum_vel}  as a function of $x_0$ and $r$. While we have tried, it has not been possible to detect any straightforward scaling relation that is satisfied in the region of interest apart from the conditions stated in the discussion above. This is due to the several timescales (resetting, relocation, diffusive), all of them of a similar order, directly competing in the process.

\section{Conclusions}

We analyzed a two-state model for a one-dimensional intermittent random walk subject to stochastic resetting at a constant rate. One state corresponds to diffusive motion, while the other represents a relocating phase where the walker moves ballistically with a constant velocity in either the right or left direction. We consider the possibility that the walker may be initially diffusing (case I) or relocating (case II). We studied the MSD and the FAT for this process first without resetting and observed that, in the unbounded domain, the MFAT diverges. 

We then studied the system with resetting and demonstrated the existence of a NESS. In addition, we computed the FATD and we proved that its mean is finite in the system with resetting and that there exists a value of the reset rate that, given all the other parameters of the model, minimizes the MFAT. Our study of the NESS, FATD, and MFAT for both cases I and II shows that they differ depending on the initial condition of the walker.  
  
We argue that this behavior occurs because, in the presence of resetting, the initial condition is consistently revisited, which prevents the smoothing out of its effect even in the long-term limit. Thus, our results demonstrate the importance of the initial condition (whether the individual starts diffusing or relocating) for the search dynamics.

Next, we conducted a numerical analysis of the MFAT, focusing on determining the parameter combination that maximizes its global optimization in the regime where ballistic and diffusive movements directly compete. For case I, we observed two distinct behaviors. There is a \textit{localized} phase in which the MFAT exhibits a well-localized minimum in $(\tau_D, \tau_v)$ phase space. On the other hand, for small $x_0$ or large $r$ there is a \textit{non-localized} phase in which a plateau region of values $(\tau_D, \tau_v)$ optimizes the MFAT. In this phase, the specific combination of $(\tau_D, \tau_v)$ is not relevant for optimizing the search. In the \textit{localized} phase, however, it is. For case II a third intermediate phase arises, \textit{ballistic localization}. Here the minimum of the MFAT is restricted to a fine line region of $(\tau_D,\tau_v)$ values, corresponding to the situation where the target can be reached on average in a single relocation. The existence of the \textit{localized phase} for a non-trivial combination of $(\tau_D,\tau_v)$ reveals that having an unresponsive relocation phase is beneficial for the optimization of the MFAT of the search process for 1D IRW in unbounded domains in the presence of resetting in agreement with \cite{BeCo05,Os07,Lo08,GoCaMe11}.

The analysis presented in this work contributes to the existing literature on search strategies and provides a foundation for future research. Recently, it has been observed that some kinds of birds combine the relocating mode with a subdiffusive detection mode \cite{O23}. Therefore, extending our IRW model in this direction in the future would be of great value. Other recent studies have focussed on the energetic cost of resetting \cite{PhysRevResearch.5.023103,olsen2023thermodynamic,PhysRevE.108.044117,Sunil_2023}, in these lines a worthwhile study of our model is to consider not only the cost of resetting but also the cost of switching phases or considering a phase more energetic than the other. In this case, the optimization of the MFAT would not suffice to characterize the efficiency of the search strategy, but what has to be optimized is the total cost of the exploration. Moreover, we note that we have presented a model with instantaneous resetting, while, for many physical and biological systems, it would be more realistic to consider it non-instantaneous such as in \cite{PhysRevE.106.044127,PhysRevE.101.052130,Pal_2019,PhysRevE.100.040101,PhysRevLett.116.170601,PhysRevE.100.042104}. In the present work we wanted to highlight the role of the trade-off between the timescales diffusion and relocation and its interplay with resetting. Thus, we chose the instantaneous resetting framework, which is more understood in the literature to put forward these results as we considered that adding a non-instantaneous reset would also introduce an extra timescale which could obscure the role of the other timescales. For future works, we aim to consider it non-instantaneous and see how these results may change, we expect a similar qualitative behavior for walkers resetting at constant speed. In a similar line, in our study we considered both the diffusive phase and the ballistic phase as part of the search process, following the approach of O. Bénichou et al in [9]. Hence, resetting can occur in either phase, restarting the search process as a whole. However, for some applications, it may be useful to modify this model so that resetting only occurs during the diffusive state and see how the optimization of the MFAT changes.

\section*{Acnowledgements}
The Authors acknowledge the financial support of the Spanish government under grant PID2021-122893NB-C22.

\appendix 

\section{Random initial condition}\label{app:randomIC}

If one considers that the walker can start in either state, diffusion or relocation, randomly then: $\bar{\rho}_{0}(k,0) = e^{-ikx_0}/2$ and $\bar{\rho}_{+}(k,0) = \bar{\rho}_{-}(k,0) = e^{-ikx_0}/4$. From Eq. \eqref{sr} the propagator for the diffusing phase $\hat{\rho}^R_0(k,s)$, reads:

\begin{equation}
    \hat{\rho}^R_0(k,s) = \frac{\hat{\rho}^D_0(k,s)}{2} + \frac{\hat{\rho}^v_0(k,s)}{2}
\end{equation}

where the superscript $R$ means random initial condition. similarly the propagator for the complete process: 

\begin{equation}
    \hat{\rho}^R(k,s) = \frac{\hat{\rho}^D(k,s)}{2} + \frac{\hat{\rho}^v(k,s)}{2}
\end{equation}

Therefore, the MSD, FAT PDF with and without resets, and steady state are simply linear combinations of cases I and II:

\begin{equation}
    \left\langle x^{2}(t)\right\rangle^R = \frac{\left\langle x^{2}(t)\right\rangle^D}{2} + \frac{\left\langle x^{2}(t)\right\rangle^v}{2}
\end{equation}

\begin{equation}
    \tilde{f}^R(x_0,s) =  \frac{\tilde{f}^D(x_0,s)}{2} + \frac{\tilde{f}^v(x_0,s)}{2}
    \label{eq:FATR}
\end{equation}

\begin{equation}
    \rho_r^{(s),R}(x) = \frac{\rho_r^{(s),D}(x)}{2} + \frac{\rho_r^{(s),v}(x)}{2}
\end{equation}

On the other hand, From Eq. \eqref{eq:MFAT-step1} and Eq. \eqref{eq:FATR} the MFAT 

\begin{align}
    \left\langle T^{R}(x_{0}) \right\rangle _r =   \frac{1}{r}\left[\frac{2(\phi_{-}-\phi_{+})}{\sigma_{-}e^{-|x_{0}|\lambda_{-}}-\sigma_{+}e^{-|x_{0}|\lambda_{+}}}-1\right]
    \label{eq:MFAT_R}
\end{align}

where $\sigma_{\pm} = \frac{\left(1+r\tau_{v}\right)\left(2+r\tau_{v}\right)}{(v\tau_{v})^{2}\lambda_{\pm}}-\lambda_{\pm}$. Then, we can see that the MFAT cannot be written as a linear combination of cases I and II. For this case, we have performed a similar parameter exploration to the one done in the section \ref{sec:parameter exploration} and we have found the same qualitative behavior of the MFAT to the one of case II. 

\section{Coefficients MSD} \label{app: Coefficients MSD}

If the walker starts at the diffusive state (case I), the coefficients of the Mean Squared Displacement, Eq. \eqref{msdsr}

\begin{align}
    A_{1}&=2\frac{\tau_v^2}{\tau_D}\frac{D-2\tau_vv^{2}-\frac{(\tau_vv)^{2}}{\tau_D}}{(1+\frac{\tau_v}{\tau_D})^{2}}, \label{msdA1d} \\
    A_{2}&=2(\tau_vv)^{2},\label{msdA2d} \\
    A_{3}&=-2\frac{\tau_v^2}{\tau_D}\frac{(D+\tau_Dv^{2})}{(1+\frac{\tau_v}{\tau_D})^{2}}.\label{msdA3d}
\end{align}

and if it is initially at the relocation state (case II),

\begin{align}
    A_{1}&=2\frac{{\tau_v}^{2}(v^{2}-\frac{D}{\tau_v})-\frac{{\tau_v}^{3} v^{2}}{\tau_D}(1+\frac{\tau_v}{\tau_D})}{(1+\frac{\tau_v}{\tau_D})^{2}},     \label{msdA1r} \\
    A_{2}&=2\tau_vv^{2}(\tau_v-\tau_D), \label{msdA2r} \\
    A_{3}&=2\tau_v\frac{( D+\tau_Dv^{2})}{(1+\frac{\tau_v}{\tau_D})^{2}}. 
    \label{msdA3r}
\end{align}

\bibliography{biblio}
\end{document}